\documentclass[pra,aps,showpacs,twocolumn,superscriptaddress]{revtex4}
\usepackage{amsfonts}
\usepackage{amsmath}
\usepackage{amssymb}
\usepackage{graphicx}
\usepackage{txfonts}

\begin{document}

\title{Quantum Fisher information of entangled coherent state in the
presence of photon losses: exact solution}
\author{Y. M. Zhang, X. W. Li}
\affiliation{Department of Physics, Beijing Jiaotong University, Beijing 100044, China}
\author{W. Yang}
\email{wenyang@csrc.ac.cn}
\affiliation{Beijing Computational Science Research Center, Beijing 100084, China}
\author{G. R. Jin}
\email{grjin@bjtu.edu.cn}
\affiliation{Department of Physics, Beijing Jiaotong University, Beijing 100044, China}
\date{\today }

\begin{abstract}
We investigate the performance of entangled coherent state for quantum
enhanced phase estimation. An exact analytical expression of quantum Fisher
information is derived to show the role of photon losses on the ultimate
phase sensitivity. We find a transition of the sensitivity from the
Heisenberg scaling to the classical scaling due to quantum decoherence of
the photon state. This quantum-classical transition is uniquely determined
by the number of photons being lost, instead of the number of incident
photons or the photon loss rate alone. Our results also reveal that a
crossover of the sensitivity between the entangled coherent state and the
NOON state can occur even for very small photon loss rate.
\end{abstract}

\maketitle

\section{Introduction}

The estimation of parameters characterizing dynamical processes is essential
to science and technology. A typical parameter estimation consists of three
steps. Firstly, the input state $|\psi _{\mathrm{in}}\rangle $ of the sensor
is prepared. Secondly, the sensor undergoes the $\phi$-dependent dynamical
process $\hat{U}(\phi )$ and evolve to the output state $|\psi\rangle$.
Finally, a measurement is made on the output state and the outcome $x$ is
used by suitable data processing to produce an unbiased estimator $\hat{\phi}(x)$ of the parameter $\phi $. The precision of the estimation is quantified by the standard deviation $\delta \phi =\langle (\hat{\phi}(x)-\phi
)^{2}\rangle $, which is determined by the input state~$|\psi _{\mathrm{in}}\rangle $~\cite{Caves,Yurke,Holland,Wineland1,Wineland2,Mitchell,Giovannetti}, the nature
of the dynamical process $\hat{U}(\phi )$~\cite{Luis,Rey,Boixo08,Choi,Woolley,Liu}, the observable being measured~\cite{Bollinger,Campos,Dowling,Kim,Lucke}, and the specific data processing
technique. The precision of the estimator $\hat{\phi}_{\mathrm{opt}}(x)$
from optimal data processing is limited by the Cram\'{e}r-Rao inequality~\cite{Helstrom,Holevo} as $\delta \phi _{\mathrm{opt}}\geq 1/\sqrt{F(\phi )}$, where $F(\phi )$ is the classical Fisher information, determined by $|\psi
_{\mathrm{in}}\rangle $, $\hat{U}(\phi )$, and the measurement scheme. Given
$|\psi _{\mathrm{in}}\rangle $ and $\hat{U}(\phi )$, maximizing $F(\phi )$
over all possible measurements gives the quantum Fisher information (QFI) $F_{Q}$ and hence the quantum Cram\'{e}r-Rao bound $\delta \phi _{\min }=1/
\sqrt{F_{Q}}$~\cite{Braunstein94,Peeze,Sun,Kacprowicz,Braunstein,Zhong} on
the attainable precision to estimate the phase $\phi$.

In general, the best precision $\delta \phi_{\min}$ improves with increasing
amount of resources $N$ employed in the measurement, e.g., the number of
photons in optical phase estimation or the total duration of measurements in
high-precision magnetic field or electric field sensing. For separable input
states, the QFI $F_{Q}\sim N$ gives the classical limit $\delta
\phi_{\min}\sim1/\sqrt{N}$, in agreement with classical central limit
theorem. To obtain an enhanced precision, it is necessary to utilize quantum
resources such as coherence, entanglement, and squeezing in the input state
for maximizing the QFI and hence the precision. This is a central issue in
quantum metrology~\cite{Paris,Ensher,XWang,Caves13}. In the absence of
noise, it has been well established that by utilizing quantum entanglement,
the QFI can be enhanced up to $F_{Q}\sim N^{2}$ and hence the precision $\delta \phi_{\min}\sim1/N$, beating the the Heisenberg limit~\cite{Wineland1,Wineland2,Mitchell,Giovannetti,Dowling2,Gerry1,Gerry2,Lee}. This
limit is ultimate estimation precision allowed by quantum resource with
definite particle number. In the presence of noises, however, it is not
clear whether the Heisenberg limit can still be achieved~\cite{Dorner,Escher2,Demkowicz}, and whether entanglement is still a useful
resource for quantum metrology.

A paradigmatic example is the estimation of relative phase shift between the
two modes propagating on different arms of the Mach-Zehnder interferometer
(MZI). Precise phase estimation is important for multiple areas of
scientific research~\cite{Dowling}, such as imaging, sensing, and information processing. In the absence of
noise, the classical limit $\delta \phi_{\min}\sim1/\sqrt{ \bar{n}}$ ($\bar{n}$ is the average number of photons) for classical coherent state can be
dramatically improved by using nonclassical states of the light. The
maximally entangled NOON states $\sim|N,0\rangle_{1,2}+|0,N \rangle_{1,2}$
(also called the GHZ state in atomic spectroscopy) has been prepared in
experiments for pursuing the Heisenberg-limited phase estimation~\cite{Wineland1,Wineland2,Mitchell}. However, the NOON states are extremely
fragile to photon losses~\cite{Enk,Dorner,Lee2,Cooper,Escher2,Joo,Demkowicz,Cooper1,Jarzyna,Knysh}. In a
lossy interferometer, it has been shown that a transition of the precision
from the Heisenberg limit to the shot-noise limit can occur with the
increase of particle number $N$~\cite{Escher2,Demkowicz}.

Recently, a specific coherent superposition of the NOON states, the
entangled coherent state (ECS) $\sim |\alpha ,0\rangle _{1,2}+|0,\alpha
\rangle _{1,2}$, was proposed as the input state for enhanced precision~\cite{Joo}. In the absence of photon losses, the precision of the ECS can surpass that of the NOON state (i.e., the Heisenberg limit, $\delta \phi _{\min }=1/\bar{n}$). In the presence of photon losses, numerical simulation suggests that the ECS outperforms the NOON state for photon numbers $\bar{n}\lesssim
5 $. For a small photon number $\bar{n}\sim 5$, the precision is better than
the classical limit by a factor $\sqrt{\bar{n}}\sim 2$. To achieve more
significant enhancement for practical applications, a much larger photon
numbers are required. The performance with a large amount of resources is an
important benchmark for a realistic quantum enhanced estimation scheme.
Therefore, a careful analysis of the QFI and the ultimate precision for the
input ECS with large $\bar{n}$ is necessary.

In this paper, we present an exact analytical result of the QFI for the
entangled coherent state with arbitrary $\bar{n}$, which provides
counter-intuitive physics that is inaccessible from previous numerical
simulations. To understand why the ECS is better than the NOON state, we first
consider an arbitrary superposition of the NOON states and find the QFI $F_{Q}\geq \bar{n}^{2}$, leading to a sub-Heisenberg limited sensitivity $\delta \phi _{\min }\leq 1/\bar{n}$. Next, we investigate the role of photon
losses on the QFI and hence the ultimate precision of the ECS. An exact
result of the QFI is derived, which is the sum of the classical term $\propto \bar{n}$ and the Heisenberg term $\propto \bar{n}^{2}$. We show that
the photon losses suppresses exponentially off-diagonal (coherence) part of
the reduced density matrix $\hat{\rho}$ and hence the Heisenberg term, while
leaving the classical term largely unchanged. The loss-induced quantum
decoherence leads to a transition of the estimation precision from the
Heisenberg scaling to the classical scaling as the number of lost photons $R\bar{n}$ increases, where $R$ is the photon loss rate and $\bar{n}$ is the mean photon number of the initial ECS. This behavior is in sharp contrast to
the NOON state, for which the photon losses eliminate completely the phase
information stored in the coherence part of $\hat{\rho}$. The ultimate
precision of the NOON state gets even worse than the classical limit when $R\bar{n}\gg 1$. Surprisingly, we find that the precision of the NOON state
may be better than that of the ECS within the crossover region at $R\bar{n}\sim 1$. This is because although the classical term of the ECS is robust
against the photon losses, the Heisenberg term decays about twice as quick
as that of the NOON state.

\section{Sub-Heisenberg limited phase sensitivity with a Superposition of
NOON states}

Firstly, let us consider an \textit{arbitrary} coherent superposition of the
NOON states as the input state after the first beam splitter of a two-mode
MZI,
\begin{equation}
\left\vert \psi _{\mathrm{in}}\right\rangle =\sum_{n=0}^{\infty }c_{n}\frac{\left\vert n\right\rangle _{1}+\left\vert n\right\rangle _{2}}{\sqrt{2}},
\label{in}
\end{equation}%
where, for brevity, we introduce the notations $|n\rangle _{1}\equiv
|n\rangle _{1}|0\rangle _{2}$ and $|n\rangle _{2}\equiv |0\rangle
_{1}|n\rangle _{2}$, representing $n$ photons in the mode $1$ (or $2$) and
the other mode in vacuum. To analyze possible achievable phase sensitivity
with $|\psi _{\mathrm{in}}\rangle $, we directly evaluate the QFI of the
outcome state after phase accumulation $|\psi (\phi )\rangle =\hat{U}(\phi
)|\psi _{\mathrm{in}}\rangle =e^{i\phi \hat{G}}|\psi _{\mathrm{in}}\rangle $, where $\hat{G}$ is the generator of phase shift. For a lossless MZI, $|\psi \rangle $ is a pure state and the QFI is given by the well-known
formula~\cite{Braunstein94,Peeze,Sun,Kacprowicz,Braunstein,Zhong}: $F_{Q}=4(\langle \psi ^{\prime }|\psi ^{\prime }\rangle -|\langle \psi
^{\prime }|\psi \rangle |^{2})=4(\langle \hat{G}^{2}\rangle -\langle \hat{G}\rangle ^{2})$, where $|\psi ^{\prime }\rangle \equiv \partial |\psi \rangle/\partial \phi $ and the expectation values are taken with respect to $|\psi
_{\mathrm{in}}\rangle $. Considering a linear phase-shift generator $\hat{G}=\hat{n}_{2}$~\cite{Dorner,Joo}, with the photon number operators $\hat{n}_{2}=\hat{a}_{2}^{\dag }\hat{a}_{2}$ and $\hat{n}_{1}=\hat{a}_{1}^{\dag}\hat{a}_{1}$, we obtain the QFI
\begin{equation}
F_{Q}=4(\langle \hat{n}_{2}^{2}\rangle -\langle \hat{n}_{2}\rangle
^{2})=2\langle \hat{n}^{2}\rangle -\langle \hat{n}\rangle ^{2},
\label{FQ-pure}
\end{equation}%
where we have used the relation $\langle \hat{n}_{1}^{l}\rangle =\langle
\hat{n}_{2}^{l}\rangle =\langle \hat{n}^{l}\rangle /2$ (for $l=1$, $2$, $\cdots$), which, together with $\langle \hat{n}_{1}\hat{n}_{2}\rangle =0$
are valid for Eq.~(\ref{in}). Since $\langle \hat{n}^{2}\rangle \geq \langle
\hat{n}\rangle ^{2}$, we have $F_{Q}\geq \bar{n}^{2}$, where $\bar{n}=\langle \hat{n}\rangle $ is the mean photon number of $|\psi _{\mathrm{in}}\rangle $. This inequality also applies to another kind of phase-shift
generator $\hat{G}=(\hat{n}_{2}-\hat{n}_{1})/2$, for which $F_{Q}=\langle
\hat{n}^{2}\rangle $. They suggest that a sub-Heisenberg limited phase
sensitivity $\delta \phi _{\min }<1/\bar{n}$ can be achievable with an arbitrary
coherent superposition of the NOON states, as Eq.~(\ref{in}). The equality $\delta \phi _{\min }=1/\bar{n}$, known as the Heisenberg limit, is attained by the NOON state~\cite{Wineland1,Wineland2,Mitchell,Giovannetti,Dowling2,Gerry1,Gerry2,Lee} $(|N\rangle _{1}+|N\rangle _{2})/\sqrt{2}$ with $\bar{n}=N$.

\begin{figure}[htpb]
\begin{centering}
\includegraphics[width=0.95\columnwidth, angle=0]{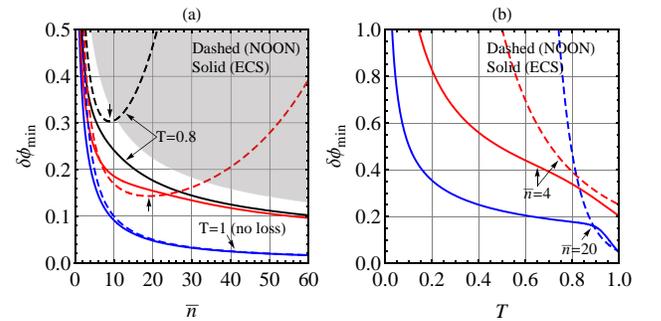}
\caption{(Color online) The ultimate precision $\delta\phi_{\min}$ against the number of photons $\bar{n}$ or $N$ (a) and the transmission rate $T$ (b) for the NOON (dashed) and the ECS (solid) states. In (a), $T=1$ (blue lines), $0.9$ (red lines), and $0.8$ (black lines). Two arrows indicate $N_{\text{opt}}=-2/\ln T$ with $T=0.8$ and $0.9$. In (b), $\bar{n}=4$ (red lines) and $20$ (blue lines). A crossover of $\delta\phi_{\min}$ between the ECS and the NOON states occurs for $\bar{n}$ (or $N$)$=20$ and $T\in(0.85, 1)$. Shaded area in (a): Region for the sensitivity worse than the shot-noise limit $1/\sqrt{\bar{n}}$.} \label{fig1}
\end{centering}
\end{figure}

Next, we review the recently proposed ECS state~\cite{Ono,Gerry}: ${\mathcal{N}}_{\alpha }(|\alpha \rangle _{1}+|\alpha \rangle _{2})$ as a special case of the superposition of NOON states, where ${\mathcal{N}}_{\alpha
}=[2(1+e^{-|\alpha |^{2}})]^{-1/2}$ is the normalization constant and $|\alpha \rangle _{1}\equiv |\alpha \rangle _{1}\vert 0\rangle
_{2} $ denotes a coherent state in the sensor mode 1 and vacuum in the
sensor mode 2 and similarly for $|\alpha \rangle _{2}\equiv |\alpha \rangle
_{2}\vert 0\rangle _{1}$. The ECS can be generated by passing a
coherent state $|\alpha /\sqrt{2}\rangle _{1}$ and a coherent state
superposition $\sim |\alpha /\sqrt{2}\rangle _{2}+|-\alpha /\sqrt{2}\rangle
_{2}$ (experimentally available for $\alpha \approx 2$~\cite{Ourjoumtsev})
through a 50:50 beam splitter~\cite{Joo}. In the absence of photon losses,
using the ECS as the input state and considering the phase accumulation
dynamics $\hat{U}(\phi )=e^{i\phi \hat{n}_{2}}$~\cite{Dorner,Joo}, we obtain
$\bar{n}=\langle \hat{n}\rangle =2{\mathcal{N}}_{\alpha }^{2}|\alpha |^{2}$,
$\langle \hat{n}^{2}\rangle =2{\mathcal{N}}_{\alpha }^{2}|\alpha
|^{2}(|\alpha |^{2}+1)$, and the quantum Fisher information
\begin{equation}
F_{Q}=2\bar{n}[1+w(\bar{n}e^{-\bar{n}})]+\bar{n}^{2},  \label{FQ_ECS0}
\end{equation}%
where we have used $\bar{n}=|\alpha |^{2}/(1+e^{-|\alpha |^{2}})$ and hence $w(\bar{n}e^{-\bar{n}})=\bar{n}e^{-|\alpha |^{2}}$. Here, $w(z)$ denotes the Lambert W function (also called the product logarithm), which gives the
principal solution for $w$ in $z=we^{w}$. For mean photon number $\bar{n}\approx |\alpha |^{2}\gg 1$, we have $w(\bar{n}e^{-\bar{n}})\approx 0$ and $
F_{Q}\approx \bar{n}(\bar{n}+2)$. From Fig.~\ref{fig1}(a), one can find that
$\delta \phi _{\min }$ of the ECS (the blue solid line) is better than that
of the NOON (the blue dashed line), especially for a modest photon number. A
recent numerical simulation shows that this improved sensitivity of the ECS
can be maintained in the presence of the photon losses for $\bar{n}\lesssim
5 $~\cite{Joo}. However, the performance of the ECS with larger $\bar{n}$
remains unclear.

\section{Quantum Fisher information and ultimate precision of Entangled
coherent state with photon losses}

In this section, we present an exact analytical expression of the QFI $F_{Q}$
and hence the ultimate precision $\delta \phi _{\min }=1/\sqrt{F_{Q}}$ for
the ECS in the presence of photon losses. This provides detailed information
for the performance of the ECS in the quantum phase estimation, especially
those at relatively large photon numbers, that are inaccessible from the
previous numerical simulation. Firstly, we derive exact analytical expression
of the quantum Fisher information for the input ECS based upon a general formula of the QFI. This formula decomposes the total QFI into three physically intuitive
contributions. Next, we present the QFI of the NOON state. Finally, by
comparing with the NOON state, we discuss the key features of the ECS and
provides a simple physics picture.

The photon losses can be modeled by inserting two identical beam splitters $\hat{B}_{k,k^{\prime }}=\exp [i(\theta /2)(\hat{a}_{k^{\prime }}^{\dag }\hat{a}_{k}+h.c.)]$ that couples two sensor modes $k=1$, $2$ and two environment
modes $k^{\prime }=1^{\prime }$, $2^{\prime }$ that are initially in the
vacuum~\cite{Enk,Dorner,Lee2,Cooper,Escher2,Joo,Demkowicz,Cooper1,Jarzyna,Knysh}. The
action of beam splitters transforms the sensor mode $\hat{a}_{k}^{\dagger}$
into a linear combination of $\hat{a}_{k}^{\dagger}$ and $\hat{a}_{k^{\prime }}^{\dagger }$: $\hat{B}_{k,k^{\prime }}\hat{a}_{k}^{\dagger }\hat{B}_{k,k^{\prime }}^{-1}=\sqrt{T}\hat{a}_{k}^{\dagger }+i\sqrt{R}\hat{a}_{k^{\prime }}^{\dagger }$, where $T=\cos ^{2}(\theta /2)$ and $R=1-T$ are transmission and absorption (loss) rates of the photons, respectively. More
specially, $T=1$ (i.e., $R=0$ ) means no photon loss and $T=0$ ($R=1$)
corresponds to complete photon loss. For the input ECS state, using $\hat{U}(\phi )|\alpha \rangle _{2}=|\alpha e^{i\phi }\rangle _{2}$ and $\hat{B}_{k,k^{\prime }}|\alpha \rangle _{k}=|\sqrt{T}\alpha \rangle _{k}|i\sqrt{R}\alpha \rangle _{k^{\prime }}$, we obtain the outcome state
\begin{eqnarray*}
|\psi (\phi )\rangle &=&{\mathcal{N}}_{\alpha }\hat{B}_{1,1^{\prime }}\hat{B}%
_{2,2^{\prime }}\hat{U}(\phi )(|\alpha \rangle _{1}+|\alpha \rangle
_{2})|0\rangle _{1^{\prime }}|0\rangle _{2^{\prime }} \\
&=&{\mathcal{N}}_{\alpha }\left( |\sqrt{T}\alpha \rangle _{1}|E^{(1)}\rangle
+|\sqrt{T}\alpha e^{i\phi }\rangle _{2}|E^{(2)}\rangle \right) ,
\end{eqnarray*}%
where the environment states are given by $|E^{(1)}\rangle \equiv |i\sqrt{R}\alpha \rangle _{1^{\prime }}|0\rangle _{2^{\prime }}$ and $|E^{(2)}\rangle\equiv |0\rangle _{1^{\prime }}|i\sqrt{R}\alpha e^{i\phi }\rangle
_{2^{\prime }}$. Tracing over them, we obtain the reduced density matrix of
the sensor modes
\begin{eqnarray}
\hat{\rho} &=&{\mathcal{N}}_{\alpha }^{2}\left\{ |\sqrt{T}\alpha \rangle
_{11}\langle \sqrt{T}\alpha |+|\sqrt{T}\alpha e^{i\phi }\rangle _{22}\langle
\sqrt{T}\alpha e^{i\phi }|\right.  \notag \\
&&\left. +\langle E^{(2)}|E^{(1)}\rangle \left( |\sqrt{T}\alpha \rangle
_{12}\langle \sqrt{T}\alpha e^{i\phi }|+h.c.\right) \right\} ,  \label{RHO}
\end{eqnarray}%
where $\langle E^{(2)}|E^{(1)}\rangle =\langle E^{(1)}|E^{(2)}\rangle
=e^{-R|\alpha |^{2}}$. Compared with the lossless case (i.e., $T=1$), the
amplitudes in the sensor modes are reduced from $|\alpha |$ to $\sqrt{T}|\alpha|$. More importantly, the photon losses suppresses the off-diagonal
coherence between the two sensor states by a factor $\langle
E^{(2)}|E^{(1)}\rangle $. We will show below that this decoherence effect
significantly degrades the estimation precision of the ECS.

Since $\hat{\rho}$ is a mixed state, to obtain the QFI one has to
diagonalize it as $\hat{\rho}=\sum_{m}\lambda _{m}|\lambda _{m}\rangle
\langle \lambda _{m}|$, where $\{|\lambda _{m}\rangle \}$ forms an
ortho-normalized and complete basis, with $\lambda _{m}$ being the weight of
$|\lambda _{m}\rangle $. According to the well-known formula~\cite%
{Braunstein94,Peeze,Sun,Kacprowicz,Braunstein,Zhong}, the QFI is given by
\begin{equation}
F_{Q}=\sum_{m,n}\frac{2}{\lambda _{m}+\lambda _{n}}\left\vert \left\langle
\lambda _{m}\right\vert \hat{\rho}^{\prime }|\lambda _{n}\rangle \right\vert
^{2},  \label{QFI-mix}
\end{equation}%
where the prime denotes the derivation about $\phi $, such as $\hat{\rho}%
^{\prime }=\partial \hat{\rho}/\partial \phi $, $\lambda _{m}^{\prime
}=\partial \lambda _{m}/\partial \phi $, and $|\lambda _{m}^{\prime }\rangle
=\partial |\lambda _{m}\rangle /\partial \phi $. Typically, the dimension of
entire Hilbert space and hence the basis $\{|\lambda _{m}\rangle\}$ is
huge, but only a small subset has nonzero weights. Therefore, using the
completeness and the ortho-normalization of $\{|\lambda _{m}\rangle\}$, we
can express the QFI in terms of the subset $\{|\lambda _{i}\rangle\}$
with $\lambda _{i}\neq 0$ (see Appendix):
\begin{equation}
F_{Q}=\sum_{i}\frac{(\lambda _{i}^{\prime })^{2}}{\lambda _{i}}%
+\sum_{i}\lambda _{i}F_{Q,i}-\sum_{i\neq j}\frac{8\lambda _{i}\lambda _{j}}{%
\lambda _{i}+\lambda _{j}}\left\vert \left\langle \lambda _{i}^{\prime
}\right\vert \lambda _{j}\rangle \right\vert ^{2},  \label{QFI-mix3}
\end{equation}%
which contains three kinds of contributions. The first term is the classical
Fisher information for the probability distribution $P(i|\phi )\equiv
\lambda _{i}(\phi )$. The second term is a weighted average over the quantum
Fisher information $F_{Q,i}=4(\langle \lambda _{i}^{\prime }|\lambda
_{i}^{\prime }\rangle -|\langle \lambda _{i}^{\prime }|\lambda _{i}\rangle
|^{2})$ for each pure state in the subset $\{|\lambda _{i}(\phi )\rangle\}$
with $\lambda _{i}\neq 0$. The last term reduces the QFI and hence the
estimation precision below the pure-state case. If the phase shift $\phi$
comes into the reduced density matrix $\hat{\rho}$ through the weights $\lambda _{i}(\phi )$ only, then the last two terms of Eq. (\ref{QFI-mix3}) give vanishing contribution to the QFI. While for $\phi$-independent weights,
however, the first term vanishes.

Compared with Eq. (\ref{QFI-mix}) that relies on the complete basis, our formula Eq. (\ref{QFI-mix3}), defined within a truncated
Hilbert space, has the advantages of faster convergence and numerical stability, especially when the reduced density matrix $\hat{\rho}$ has some
eigenvectors with extremely small but nonvanishing weights.

For the input ECS, we note that the reduced density matrix $\hat{\rho}$ only
contains two sensor states $|\sqrt{T}\alpha \rangle _{1}$ and $|\sqrt{T}\alpha e^{i\phi }\rangle _{2}$ [see Eq.~(\ref{RHO})]. This feature enables
us to expand $\hat{\rho}$ in terms of two eigenvectors with nonzero
eigenvalues (see Append. A),
\begin{equation}
\hat{\rho}=\lambda _{+}\left\vert \lambda _{+}(\phi )\right\rangle \langle
\lambda _{+}(\phi )|+\lambda _{-}\left\vert \lambda _{-}(\phi )\right\rangle
\langle \lambda _{-}(\phi )|,  \label{diag}
\end{equation}%
where the eigenvalues $\lambda _{\pm }={\mathcal{N}}_{\alpha }^{2}(1\pm
e^{-R|\alpha |^{2}})(1\pm e^{-T|\alpha |^{2}})$ are $\phi $-independent and
obey $\lambda _{-}+\lambda _{+}=1$. The phase-dependent eigenvectors are
given by
\begin{equation}
|\lambda _{\pm }(\phi )\rangle =\eta _{\pm }\left[ \pm |\sqrt{T}\alpha
\rangle _{1}+|\sqrt{T}\alpha e^{i\phi }\rangle _{2}\right] ,
\label{eigenvectors}
\end{equation}%
with the normalization factors $\eta _{\pm }=1/\sqrt{2(1\pm e^{-T|\alpha
|^{2}})}$. It is easy to prove that $\langle \lambda _{\pm }|\lambda _{\pm
}\rangle =1$ and $\langle \lambda _{+}|\lambda _{-}\rangle =\langle \lambda
_{+}|\hat{\rho}|\lambda _{-}\rangle =0$. Using Eq.~(\ref{QFI-mix3}), we
obtain exact analytical expression of the QFI (see Append. B):
\begin{equation}
F_{Q}=F_{Q}^{\mathrm{cl}}+F_{Q}^{\mathrm{HL}},  \label{FQ_ECS}
\end{equation}%
where the classical term $F_{Q}^{\mathrm{cl}}=2\bar{n}T[1+Tw(\bar{n}e^{-\bar{n}})]$, and the Heisenberg term
\begin{equation}
F_{Q}^{\mathrm{HL}}=(\bar{n}T)^{2}\left( \frac{e^{-2R|\alpha
|^{2}}-e^{-2T|\alpha |^{2}}}{1-e^{-2T|\alpha |^{2}}}\right) .  \label{FQHL}
\end{equation}%
In the absence of photon losses (i.e., $R=0$ and $T=1$), our result reduces
to the lossless case, i.e., Eq.~(\ref{FQ_ECS0}). Compared with it, we find
that the photon losses leads to two effects on the QFI (and
hence the estimation precision). Firstly, it trivially reduces the photon
number from $\bar{n}$ in the input state to $\bar{n}T$ in the output state.
Secondly, it exponentially suppresses the QFI from $F_{Q}^{\mathrm{HL}}\sim(\bar{n}T)^{2}$ to the classical scaling $\sim 2\bar{n}T$ (see below).

For a comparison, we also employ Eq. (\ref{QFI-mix3}) to derive the QFI for
the NOON state $(|N\rangle _{1}+|N\rangle _{2})/\sqrt{2}$. It is easy to
write down the reduced density matrix in a diagonal form:
\begin{equation}
\hat{\rho}=\sum_{n=0}^{N-1}\lambda _{n}(\left\vert n\right\rangle
_{11}\left\langle n\right\vert +\left\vert n\right\rangle _{22}\left\langle
n\right\vert )+T^{N}\left\vert \psi _{\mathrm{NOON}}\right\rangle
\left\langle \psi _{\mathrm{NOON}}\right\vert ,  \label{N00N}
\end{equation}%
where the first part is an incoherent mixture of Fock states $|n\rangle _{1}$
and $|n\rangle _{2}$ with $\phi $-independent weights $\lambda _{n}=\tbinom{N}{n}T^{n}R^{N-n}/2$. The phase information is stored in the second part, $|\psi _{\mathrm{NOON}}\rangle =(|N\rangle _{1}+e^{iN\phi }|N\rangle _{2})/\sqrt{2}$, which, for the lossless case, gives the QFI $N^{2}$. Therefore,
according to Eq.~(\ref{QFI-mix3}), the total QFI is equal to the QFI\ of $|\psi _{\mathrm{NOON}}\rangle $ times its weight $T^{N}$, namely
\begin{equation}
F_{Q,\mathrm{NOON}}=N^{2}T^{N},  \label{FQ_NOON}
\end{equation}%
in agreement with Ref.~\onlinecite{Dorner}. With increasing photon number $N$, the ultimate precision $\delta \phi _{\min }=T^{-N/2}/N$ shows a global minimum at $N_{\mathrm{opt}}=-2/\ln T\approx 2/R$ (as $T=1-R\approx e^{-R}$
for small $R$), indicated by the arrows of Fig.~\ref{fig1}(a).

In Fig.~\ref{fig1}(a), we plot $\delta \phi_{\min}$ of the ECS (the NOON)
state as a function of number of photons $\bar{n}$ ($N$) for the
transmission rates $T=0.8$, $0.9$, and $1$ (from top to bottom). Regardless
of $T$, one can find that $\delta \phi_{\min}$ of the input ECS decreases
monotonically with the increase of $\bar{n}$. While for the NOON state,
however, $\delta \phi_{\min}$ reaches its minimum at $N_{\mathrm{opt}}$ and
then grows rapidly. In Fig.~\ref{fig1}(b), we show $\delta \phi_{\min}$
against $T$ for $\bar{n}$ ($N$)$=4$ and $20$. It is interesting to note that
a crossover of $\delta \phi_{\min}$ between the ECS and the NOON states
occurs for large $\bar{n}$ and $T$ (say, $T>0.85$).

\begin{figure}[hptb]
\includegraphics[width=\columnwidth, angle=0]{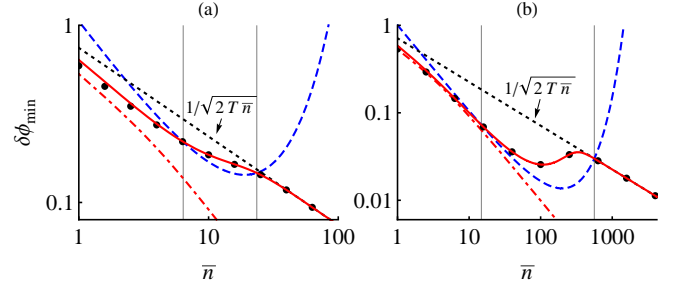}
\caption{(Color online) log-log plot of $\delta\phi_{\min}$
for $T=0.9$ (a) and $T=0.99$ (b). Black dotted line: the classical limit $1/\sqrt{2T\bar{n}}$; Blue dashed line: $\delta\phi_{\min}$ of the NOON state; Red solid line (circles): approximated (exact) $\delta\phi_{\min}$ of the ECS; Red dot-dashed line: $\delta\phi_{\min}$ of the ECS in the absence of photon losses (i.e, $T=1$), given by Eq.~(\ref{FQ_ECS0}). The two vertical lines at $\bar{n}=6.4$ and $23.5$ in (a) and $\bar{n}=14.8$ and $561$ in (b) show the crossover of $\delta\phi_{\min}$ between the ECS and the NOON states.}
\label{fig2}
\end{figure}

We now analyze the QFI under practical conditions: $T\sim 1$ ($R\sim 0$) and
$|\alpha |^{2}\gg 1$, for which $w(\bar{n}e^{-\bar{n}})\approx 0$ and hence $|\alpha |^{2}\approx \bar{n}$. In addition, the exponential term $e^{-2T|\alpha |^{2}}$ of Eq.~(\ref{FQHL}) is negligible. As a result, the
QFI reduces to
\begin{equation}
F_{Q}=F_{Q}^{\mathrm{cl}}+F_{Q}^{\mathrm{HL}}\approx 2\bar{n}T+(\bar{n}%
T)^{2}e^{-2R\bar{n}},  \label{FQ_ECS_APPROX}
\end{equation}%
where the exponential term $e^{-2R\bar{n}}=|\langle E^{(2)}|E^{(1)}\rangle
|^{2}$, quantifies the off-diagonal coherence of the sensor states. When the
number of photons being lost $R\bar{n}=(1-T)\bar{n}\ll 1$, the Heisenberg
term $F_{Q}^{\mathrm{HL}}\approx (\bar{n}T)^{2}e^{-2R\bar{n}}$ dominates and
the ultimate precision obeys $\delta \phi _{\min }^{\mathrm{HL}}\approx e^{R\bar{n}}/(\bar{n}T)$. With the increase of $R\bar{n}$, the classical term $F_{Q}^{\mathrm{cl}}\approx 2\bar{n}T$ becomes important. As $R\bar{n}\gg 1$, a complete decoherence of the two sensor states occurs due to $|\langle
E^{(2)}|E^{(1)}\rangle |^{2}\rightarrow 0$, leading to the completely mixed
state
\begin{equation}
\hat{\rho}\approx \frac{1}{2}(|\sqrt{T}\alpha \rangle _{11}\langle \sqrt{T}%
\alpha |+|\sqrt{T}\alpha e^{i\phi }\rangle _{22}\langle \sqrt{T}\alpha
e^{i\phi }|),  \label{ECS_RHO}
\end{equation}%
where the first term $|\sqrt{T}\alpha \rangle _{11}\langle \sqrt{T}\alpha |$
carries no phase information and hence $F_{Q,1}=0$, and the $\phi$-dependent second term $|\sqrt{T}\alpha e^{i\phi }\rangle _{22}\langle \sqrt{T}\alpha e^{i\phi }|$ produces the pure-state QFI $F_{Q,2}\approx 4\bar{n}T$. Therefore, according to Eq.~(\ref{QFI-mix3}), the total QFI of $\hat{\rho}$ reads $F_{Q}\approx \sum_{i}\lambda _{i}F_{Q,i}\approx 2\bar{n}T$, which in
turn gives the classical scaling of the sensitivity $\delta \phi _{\min
}\approx \delta \phi _{\min }^{\mathrm{cl}}\approx 1/\sqrt{2\bar{n}T}$. In
Fig.~\ref{fig2}, we present the log-log plot of $\delta \phi _{\min }$ for
the loss rate $R=0.1$ and $0.01$. As shown by the red solid line, the simple
formula of Eq.~(\ref{FQ_ECS_APPROX}) agrees quite well with the exact result
(the solid circles). They both show a turning point at $\bar{n}\sim 1/R$.
Indeed, the crossover of the quantum-classical transition takes place when
the Heisenberg term $F_{Q}^{\mathrm{HL}}$ is comparable with the classical
term $F_{Q}^{\mathrm{cl}}$, i.e., $R\bar{n}\sim 1$.

For the ECS with large photon losses, i.e., $R\bar{n}\gg 1$, the ultimate
precision $\delta \phi _{\min }$ obeys the classical scaling $1/\sqrt{2\bar{n}T}$, which is conformed by Fig.~\ref{fig2}. The precision of the NOON state is optimal at $\bar{n}=-2/\ln T\approx 2/R$ and then rapidly degrades below
the classical limit [see the dashed lines, also Fig.~\ref{fig1}(a)]. This is
in sharp contrast to the ECS state. Qualitatively, different behaviors of
the two states arises from the different influences of photon losses:

\begin{enumerate}
\item For the ECS $\sim |\alpha \rangle _{1}+|\alpha \rangle _{2}$, the
off-diagonal coherence between the two sensor states $|\sqrt{T}\alpha
\rangle _{1}$ and $|\sqrt{T}\alpha e^{i\phi }\rangle _{2}$ is exponentially
suppressed by the photon losses, but the diagonal components of $\hat{\rho}$
still carries the phase information [see Eq.~(\ref{ECS_RHO})], which
contributes the QFI $F_{Q}\approx 2\bar{n}T$.

\item For the NOON state $\sim |N\rangle _{1}+|N\rangle _{2}$, the phase
information is stored only in the coherence part of $\hat{\rho}$ [see Eq.~(\ref{N00N})], which decays with the photon losses as $T^{N}\approx e^{-RN}$ for small $R$. When the lost photon number $RN\gg 1$, the information about
the phase shift $\phi $ is completely eliminated.
\end{enumerate}

From Fig.~\ref{fig1}, we have observed the crossover of $\delta \phi _{\min}$ between the ECS and the NOON states, which can be understood by simply
comparing the QFIs for the two states. Without the photon losses, the
ultimate precision of the ECS always surpass those of the NOON states
because $F_{Q}=F_{Q}^{\mathrm{cl}}+F_{Q}^{\mathrm{HL}}>F_{Q,\mathrm{NOON}}$
(as $F_{Q}^{\mathrm{HL}}=F_{Q,\mathrm{NOON}}=\bar{n}^{2}$). In the presence
of moderate photon losses, the Heisenberg term $F_{Q}^{\mathrm{HL}}\approx (\bar{n}T)^{2}e^{-2R\bar{n}}$ decays more quickly than that of the NOON state $F_{Q,\mathrm{NOON}}\approx \bar{n}^{2}e^{-R\bar{n}}$. This makes it possible for the NOON\ state to outperform the ECS when the quantum contribution $F_{Q}^{\mathrm{HL}}$ dominates the classical
contribution $F_{Q}^{\mathrm{cl}}$. From Fig.~\ref{fig2}, one can find that
the NOON states with $\bar{n}\in (6.4$, $23.5)$ for $R=0.1$ and $\bar{n}\in(14.8$, $561)$ for $R=0.01$ are preferable, within the vertical lines of Fig.~\ref{fig2}.

\begin{figure}[ptb]
\includegraphics[width=\columnwidth]{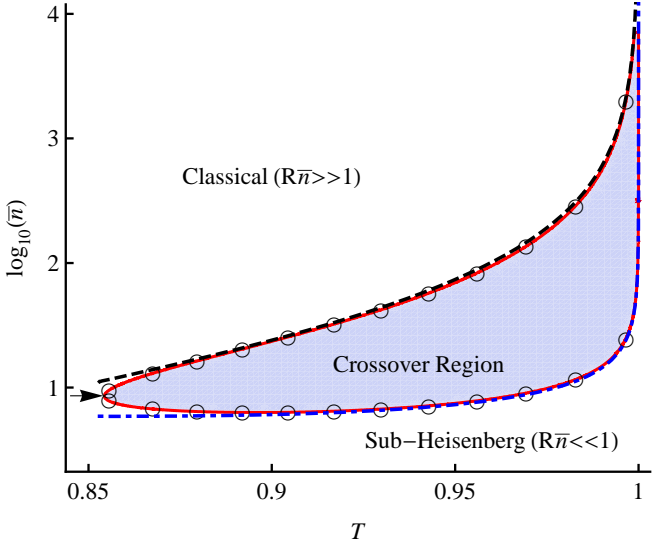}
\caption{(Color online) The crossover region in which $\delta\phi_{\min}$ of the NOON state is preferable. Red solid line: $\bar{n}T^{
\bar{n}-1}=2+\bar{n}Te^{-2R\bar{n}}$ for $R=1-T$ and $T\in(0.85,1)$; Black
dashed and blue dot-dashed lines: $\bar{n}_{u}\simeq 3.2T^{6}/R^{1.15}$ and $\bar{ n}_{l}\simeq1.4T^{-3}/R^{0.5}$, fitting very well with the boundary of the crossover region (Open circles). The critical point of the crossover: ($T$, $\bar{n}$ )=($0.854$, $8.58$), as indicated by the arrow.}
\label{fig3}
\end{figure}

In general, the crossover condition can be obtained by equating Eq.~(\ref{FQ_ECS_APPROX}) and Eq.~(\ref{FQ_NOON}). This gives a transcendental
equation: $\bar{n}T^{\bar{n}-1}\approx 2+\bar{n}Te^{-2R\bar{n}}$, as
illuminated by the red solid curve in Fig.~\ref{fig3}. It shows that the
NOON states outperforms the ECS inside the crossover region, while the ECS
prevails outside. The upper and the lower boundaries of the region are well
fitted by $\bar{n}_{u}\approx 3.2T^{6}/R^{1.15}$ (the black dashed line) and
$\bar{n}_{l}\approx 1.4T^{-3}/R^{1/2}$ (the blue dash-dotted line),
respectively. The upper boundary corresponds to $F_{Q,\mathrm{NOON}}\approx
F_{Q}^{\mathrm{cl}}$. As shown in Fig.~\ref{fig3}, we find that the
crossover of $\delta \phi _{\min }$ between the ECS and the NOON states
takes place for $T\in (0.854,1)$. For such a relatively low loss rate ($0<R<0.15$), the precision of the NOON state could surpass that of the ECS
over a wider range of $\bar{n}$ until the classical term $F_{Q}^{\mathrm{cl}}$ begins to dominate. However, the NOON states with $\bar{n}>\bar{n}_{u}$
ceases to be optimal and its precision gets even worse than the classical
limit~\cite{Dorner}. From Fig.~\ref{fig3}, we also note that no crossover
occurs for $T\lesssim 0.854$ and the ultimate precision of the ECS is always
better than that of the NOON state [see also the black lines of Fig.~\ref{fig1}(a)].

\section{Conclusion}

By considering a superposition of NOON states as \textquotedblleft input" of
a lossless optical interferometer, we have shown that the quantum Fisher
information $F_{Q}\geq \bar{n}^{2}$ and therefore the ultimate precision of
phase sensitivity can be better than the Heisenberg limit.

As a special case of the superposed state, an entangled coherent state $\propto |\alpha ,0\rangle _{1,2}+|0,\alpha \rangle _{1,2}$ has been
investigated. Exact result of the quantum Fisher information is obtained to
investigate the role of photon losses on the lower bound of phase
sensitivity $\delta \phi _{\min }$. Without the photon losses, i.e., the
absorption rate $R=0$ and the transmission rate $T=1$, we confirm that the
input ECS always outperform the NOON state~\cite{Joo}. In the presence of
photon losses, the transition of $\delta \phi _{\min }$ from the Heisenberg
scaling to the classical limit occurs due to the loss-induced quantum
decoherence between the sensor states. The quantum-classical transition
depends upon the number of photons being lost $R\bar{n}$, but rather the
total photon number $\bar{n}$ or the lose rate $R$ alone. For a given
transmission rate $T\in (0.85, 1)$, we also find that there exists a
crossover of $\delta \phi _{\min}$ between the ECS and the NOON states. The
NOON state is preferable in the crossover region, i.e., $\bar{n}T^{\bar{n}-1}\gtrsim 2+\bar{n}Te^{-2R\bar{n}}$. For $R\bar{n}\gg 1$, however, the
precision of the NOON state degrades below the classical limit; While for
the ECS state, $\delta \phi _{\min }$ obeys the classical limit $1/\sqrt{2T\bar{n}}$, better than that of the NOON state.

\begin{acknowledgments}
We thank Professor D. L. Zhou and Professor J. P. Dowling for helpful
discussions. This work is supported by Natural Science Foundation of China
(NSFC, Contract Nos.~11174028 and ~11274036), the Fundamental Research Funds
for the Central Universities (Contract No.~2011JBZ013), and the Program for
New Century Excellent Talents in University (Contract No.~NCET-11-0564).
X.W.L is partially supported by National Innovation Experiment Program for
University Students.
\end{acknowledgments}

\appendix

\section{Eigenvalues and Eigenvectors of the reduced density matrix}

We present a general method to diagonalize the reduced density matrix likes
Eq.~(\ref{RHO}). The eigenvector of $\hat{\rho}$ can be spanned as $|\lambda(\phi)\rangle=\sum_{j}c_{j}|\Phi_{j}\rangle$, where the states $|\Phi_{j}\rangle$ are not necessary orthogonal. Using the eigenvalue
equation: $\hat{\rho}|\lambda(\phi)\rangle=\lambda|\lambda(\phi)\rangle$, or
equivalently $\sum_{j}\langle \Phi_{i}|\hat{\rho}|\Phi_{j}\rangle
c_{j}=\lambda \sum_{j}\langle \Phi_{i}|\Phi_{j}\rangle c_{j} $, we can
determine the eigenvalue $\lambda$ and the amplitudes $c_{j}$. It is
convenient to write down the eigenvalue equation in a matrix form: $\boldsymbol{\rho}\mathbf{c}=\lambda \mathbf{Ac}$, where the elements of $\boldsymbol{\rho}$ and $\mathbf{A}$ are $\rho_{ij}=\langle \Phi_{i}|\hat{\rho }|\Phi_{j}\rangle$ and $A_{ij}=\langle \Phi_{i}|\Phi_{j}\rangle$, and $\mathbf{c}=(c_{1},c_{2},\cdots)^{T}$. Multiplying the inverse matrix $\mathbf{A}^{-1}$ on the left, we can rewrite the eigenvalue equation as
\begin{equation}
\boldsymbol{\tilde{\rho}}\mathbf{c}\equiv \mathbf{A}^{-1}\boldsymbol{\rho}
\mathbf{c}=\lambda \mathbf{c},  \label{eigen eq}
\end{equation}
where $\boldsymbol{\tilde{\rho}}=\mathbf{A}^{-1}\boldsymbol{\rho}$.

Using the above formula, we now diagonalize the reduced density operator of
Eq.~(\ref{RHO}). Firstly, we expand the eigenvectors as $|\lambda (\phi
)\rangle =c_{1}|\Phi_{1}\rangle +c_{2}|\Phi_{2}\rangle $, where $|\Phi
_{1}\rangle =|\sqrt{T}\alpha \rangle _{1}=|\sqrt{T}\alpha \rangle
_{1}|0\rangle_{2}$ and $|\Phi_{2}\rangle =|\sqrt{T}\alpha e^{i\phi
}\rangle_{2}=|0\rangle_{1}|\sqrt{T}\alpha e^{i\phi}\rangle_{2}$. It is easy
to obtain the matrix
\begin{equation*}
\boldsymbol{\tilde{\rho}}={\mathcal{N}}_{\alpha }^{2}\left(
\begin{array}{cc}
1+e^{-|\alpha |^{2}} & e^{-T|\alpha |^{2}}+e^{-R|\alpha |^{2}} \\
e^{-T|\alpha |^{2}}+e^{-R|\alpha |^{2}} & 1+e^{-|\alpha |^{2}}%
\end{array}
\right) ,
\end{equation*}
where $T$ ($R=1-T$) are the transmission (absorption) rate of the photons
and ${\mathcal{N}}_{\alpha}^2=1/[2(1+e^{-|\alpha |^{2}})]$. Next, from the
equation $|\lambda \mathbf{I}-\boldsymbol{\tilde{\rho}}|=0$, we obtain the
eigenvalues
\begin{equation}
\lambda _{\pm }={\mathcal{N}}_{\alpha}^{2}\left[ \left( 1+e^{-|\alpha
|^{2}}\right) \pm \left( e^{-T|\alpha |^{2}}+e^{-R|\alpha |^{2}}\right)\right] ,  \label{lambda}
\end{equation}
which obeys $\lambda_{-}+\lambda_{+}=1$. Substituting $\lambda_{\pm}$ into
Eq.~(\ref{eigen eq}), or $(\lambda \mathbf{I}-\boldsymbol{\tilde{\rho}})
\mathbf{c}=0$, we further obtain the amplitudes $c_{1}=\pm c_{2}$, i.e., the
eigenvectors $|\lambda_{\pm}(\phi)\rangle \varpropto (\pm |\sqrt{T} \alpha
\rangle_{1}+|\sqrt{T}\alpha e^{i\phi}\rangle_{2}) $, as Eq.~(\ref{eigenvectors}).

\section{Derivations of the quantum Fisher information}

Firstly, we derive the general expression of the QFI [i.e., Eq.~(\ref%
{QFI-mix3})]. For a mixed state $\hat{\rho}=\sum_{m}\lambda _{m}|\lambda
_{m}\rangle \langle \lambda _{m}|$, the QFI is given by the well-known
formula of Eq.~(\ref{QFI-mix}), where the eigenvectors of the reduced density matrix
$\{|\lambda _{m}\rangle\}$ span an otho-normalized and complete basis. In
general, the dimension of the entire Hilbert space is huge. However, there
exists a much smaller subset $\{|\lambda _{i}\rangle\}$ with nonzero
weights $\lambda _{i}$. It is convenient to express the QFI in terms of this
subset only. For this purpose, we divide the complete basis $\{|\lambda _{m}\rangle \}$ into two subsets: $\{\vert \lambda _{i}\rangle \}$ and $\{\vert\lambda _{\bar{\imath}}\rangle\}$, with $\lambda _{i}\neq 0$ and $\lambda _{\bar{\imath}}=0$,
respectively. Using the completeness relation $\sum_{\bar{\imath}}\vert
\lambda _{\bar{\imath}}\rangle \langle \lambda _{\bar{\imath}}\vert
=1-\sum_{i}\vert \lambda _{i}\rangle\langle\lambda_{i}\vert$, Eq. (\ref{QFI-mix}) can be rewritten as%
\begin{eqnarray}
F_{Q}& =&\sum_{i}\frac{2\langle \lambda _{i}\vert (\hat{\rho}^{\prime
})^{2}|\lambda _{i}\rangle }{\lambda _{i}}+\sum_{j}\frac{2\langle \lambda
_{j}\vert (\hat{\rho}^{\prime })^{2}|\lambda _{j}\rangle }{\lambda _{j}}
\notag \\
&& +\sum_{i,j}2\left( \frac{1}{\lambda _{i}+\lambda _{j}}-\frac{1}{\lambda
_{i}}-\frac{1}{\lambda _{j}}\right) \left\vert \left\langle \lambda
_{i}\right\vert \hat{\rho}^{\prime }|\lambda _{j}\rangle \right\vert ^{2},
\label{QFI4}
\end{eqnarray}%
where only the subset $\{|\lambda _{i}\rangle \}$ with $\lambda _{i}\neq 0$
is involved. Since $\{|\lambda _{i}\rangle \}$ are ortho-normalized, i.e., $\langle \lambda _{i}|\lambda _{j}\rangle =\delta _{i,j}$, we have $\langle\lambda _{i}|\lambda _{j}^{\prime }\rangle +\langle \lambda _{i}^{\prime}|\lambda _{j}\rangle =0$, and hence
\begin{eqnarray*}
\langle \lambda _{i}|(\hat{\rho}^{\prime })^{2}|\lambda _{i}\rangle &=&(
\lambda _{i}^{\prime }) ^{2}+\lambda _{i}^{2}\langle \lambda _{i}^{\prime
}|\lambda _{i}^{\prime }\rangle \\
&&+\sum_{l}( \lambda _{l}^{2}-2\lambda _{i}\lambda _{l}) |\langle \lambda
_{i}^{\prime }|\lambda _{l}\rangle |^{2}, \\
\vert \langle \lambda _{i}\vert \hat{\rho}^{\prime }\vert \lambda
_{j}\rangle \vert ^{2} &=&(\lambda _{i}^{\prime })^{2}\delta _{i,j}+(\lambda
_{i}-\lambda _{j})^{2}|\langle \lambda _{i}^{\prime }|\lambda _{j}\rangle
|^{2}.
\end{eqnarray*}%
Substituting them into Eq.~(\ref{QFI4}), and using $|\langle \lambda
_{j}^{\prime }|\lambda _{i}\rangle |^{2}=|\langle \lambda _{i}^{\prime
}|\lambda _{j}\rangle |^{2}$, we obtain the general formula of the QFI as
main text of Eq.~(\ref{QFI-mix3}).

Now we apply the general formula to calculate the QFI of the ECS state.
Since the eigenvalues of the reduced density matrix $\lambda_{\pm}$ are
phase-independent, the first term of Eq.~(\ref{QFI-mix3}) vanishes. From
Eq.~(\ref{eigenvectors}), it is easy to obtain the derivation of the
eigenvectors
\begin{equation}
|\lambda _{\pm }^{\prime }\rangle =\eta _{\pm }\frac{\partial }{\partial
\phi }|\sqrt{T}\alpha e^{i\phi }\rangle _{2}=\eta _{\pm }\sum_{n=0}^{\infty
}ind_{n}(\alpha \sqrt{T}e^{i\phi })\left\vert n\right\rangle _{2},
\label{derivation}
\end{equation}%
where the normalization factors $\eta _{\pm }=1/\sqrt{2(1\pm e^{-T|\alpha
|^{2}})}$ and the probability amplitudes of coherent state $d_{n}(\alpha
)\equiv \langle n|\alpha \rangle =\alpha ^{n}e^{-\frac{1}{2}|\alpha |^{2}}/\sqrt{n!}$, which satisfiy\ $\sum_{n}|d_{n}(\alpha )|^{2}=1$ and
\begin{equation*}
\sum_{n=0}^{+\infty }n\left\vert d_{n}(\alpha )\right\vert ^{2}=|\alpha |^{2}
\text{, \ }\sum_{n=0}^{+\infty }n^{2}\left\vert d_{n}(\alpha )\right\vert
^{2}=|\alpha |^{2}(1+|\alpha |^{2}).
\end{equation*}%
Therefore, combining Eq.~(\ref{eigenvectors}) and Eq.~(\ref{derivation}), we
obtain
\begin{equation}
\langle \lambda _{\pm }|\lambda _{\pm }^{\prime }\rangle =\eta _{\pm
}^{2}\sum_{n=0}^{\infty }in\left\vert d_{n}(\alpha e^{i\phi }\sqrt{T}
)\right\vert ^{2}=i\eta _{\pm }^{2}|\alpha |^{2}T,  \label{innerP1}
\end{equation}%
and similarly, $\langle \lambda _{\mp }|\lambda _{\pm }^{\prime }\rangle
=i\eta _{+}\eta _{-}|\alpha |^{2}T$, as well as $\langle \lambda _{\pm
}^{\prime }|\lambda _{\pm }^{\prime }\rangle =\eta _{\pm }^{2}|\alpha
|^{2}T(1+|\alpha |^{2}T)$. These results enable us to calculate the
remaining terms of Eq. (\ref{QFI-mix3}):
\begin{eqnarray}
\sum_{i=\pm }\lambda _{i}F_{Q,i} &=&4\lambda _{+}\eta _{+}^{2}|\alpha
|^{2}T(1+|\alpha |^{2}T-\eta _{+}^{2}|\alpha |^{2}T)  \notag \\
&&+4\lambda _{-}\eta _{-}^{2}|\alpha |^{2}T(1+|\alpha |^{2}T-\eta
_{-}^{2}|\alpha |^{2}T),  \label{t2}
\end{eqnarray}%
and
\begin{equation}
\sum_{i=\pm ,j=\mp }\frac{8\lambda _{i}\lambda _{j}}{\lambda _{i}+\lambda
_{j}}\left\vert \left\langle \lambda _{i}^{\prime }\right\vert \lambda
_{j}\rangle \right\vert ^{2}=16\lambda _{+}\lambda _{-}\eta _{+}^{2}\eta
_{-}^{2}|\alpha |^{4}T^{2},  \label{t3}
\end{equation}%
due to $\lambda _{+}+\lambda _{-}=1$. Finally, we get the exact result of
the QFI for the input ECS:
\begin{equation*}
F_{Q}=4{\mathcal{N}}_{\alpha }^{2}|\alpha |^{2}T\left[ 1+|\alpha |^{2}T-{\
\mathcal{N}}_{\alpha }^{2}|\alpha |^{2}T\left( 1+\frac{1-e^{-2R|\alpha |^{2}}}{1-e^{-2T|\alpha |^{2}}}\right) \right] ,
\end{equation*}%
where we have used the relations: $\lambda _{+}\eta _{+}^{2}+\lambda
_{-}\eta _{-}^{2}={\mathcal{N}}_{\alpha }^{2}$ and $4\lambda _{+}\lambda
_{-}\eta _{+}^{2}\eta _{-}^{2}={\mathcal{N}}_{\alpha }^{4}[1-e^{-2R|\alpha
|^{2}}]$, as well as
\begin{equation*}
\lambda _{+}\eta _{+}^{4}+\lambda _{-}\eta _{-}^{4}=\frac{{\mathcal{N}}%
_{\alpha }^{2}}{2}\frac{1-e^{-|\alpha |^{2}}}{1-e^{-2T|\alpha |^{2}}}.
\end{equation*}%
Using $\bar{n}=2{\mathcal{N}}_{\alpha }^{2}|\alpha |^{2}$ and hence $|\alpha|^{2}=\bar{n}+w(\bar{n}e^{-\bar{n}})$, the QFI can be further simplified as Eq.~(\ref{FQ_ECS}).


\end{document}